\documentclass[a4paper]{article} 	% use "amsart" instead of "article" for AMSLaTeX format
\usepackage{geometry}  		% See geometry.pdf to learn the layout options. There are lots.
\usepackage{graphicx,subcaption}					
\usepackage{amssymb}
\usepackage{indentfirst}
\usepackage{amsmath}
\usepackage{amsthm}
\usepackage{bm}
\usepackage{lineno}
\usepackage{setspace}
\usepackage{booktabs,multirow}
\usepackage{authblk}
\usepackage{subcaption}
\usepackage{graphicx}
\usepackage{float}
\usepackage[flushleft]{threeparttable} % For adding footnotes to tables

\RequirePackage[colorlinks,citecolor=blue,urlcolor=blue]{hyperref}

\newcommand{\tr}{\mathrm{tr}}
\newcommand{\R}{\texttt{R}}

\newcommand{\Matern}{Mat\'ern }
\newcommand{\N}{\mathcal{N}}
\newcommand{\AR}{\mathrm{AR1}}

\usepackage[url=false, isbn=false, eprint=false, backref=true, style= authoryear, backend=bibtex, maxcitenames=2, giveninits=true, maxbibnames=100]{biblatex}
\DeclareNameAlias{sortname}{last-first}
\title{Optimal design for on-farm strip trials --- systematic or randomised?}
\author[1]{Zhanglong Cao}
\author[1]{Andrew Grose}
\author[1]{Jordan Brown}
\author[1,2]{Suman Rakshit}

\affil[1]{SAGI West, School of Molecular and Life Sciences, Curtin University, Perth, Australia}
\affil[2]{School of Electrical Engineering, Computing and Mathematical Sciences, Curtin University, Perth, Australia}
 
\date{}							
\begin{document}

\maketitle
	
\begin{abstract}
There is no doubt on the importance of randomisation in agricultural experiments by agronomists and biometricians. Even when agronomists extend the experimentation from small trials to large on-farm trials, randomised designs predominate over systematic designs. However, the situation may change depending on the objective of the on-farm experiments (OFE). If the goal of OFE is obtaining a smooth map showing the optimal level of a controllable input across a grid made by rows and columns covering the whole field, a systematic design should be preferred over a randomised design in terms of robustness and reliability. With the novel geographically weighted regression (GWR) for OFE and simulation studies, we conclude that, for large OFE strip trials, the difference between randomised designs and systematic designs are not significant if a linear model of treatments is fitted or if spatial variation is not taken into account. But for a quadratic model, systematic designs are superior to randomised designs. 
\end{abstract}
	

\section{Introduction}\label{Sec:Intro}

The principles of randomisation were first expounded in 1925, when \textcite{Fisher1934Statistical} analysed a few systematically arranged experiments and pointed out that randomisation can provide valid tests of significance subject to appropriate restrictions, such as experimental units arranged in blocks or in the rows and columns of a Latin square \parencite{Verdooren2020History}. The most straightforward, least restrictive experimental design is the completely randomised design (CRD). More complex designs, such as randomised complete block design, split-plot design, strip-plot design and latin square design, are also widely used in agricultural experiments \parencite{Petersen1994Agricultural}, Following these principles, randomised designs are routinely used from small on-site trials to large on-farm experiments (OFE) in contrast to systematic designs, which are rarely used. 

% Old concluding sentence: With the increase in complexity, randomised block design, split-plot design, strip-plot design and Latin square design are all widely used in agricultural experiments \parencite{Petersen1994Agricultural}. 

OFE enables farmers the flexibility to implement large-scale experiments in order to test management practices on their farms \parencite{Evans2020Assessment}. The aim of OFE is to enable farmers to improve their competence of uncertainties and to take into account their existing strengths of handling translational and structural uncertainty \parencite{Cook2013Onfarm}. In the situation that the goal is to compare yield responses between management classes or to select individuals with the best performance as new market varieties, a randomised design is superior to a systematic design \parencite{Pringle2004FieldScale, Selle2019Flexible}. 

%Like conventional agricultural experiments, 

Randomisation has been considered a crucial prerequisite for obtaining valid statistical inferences \parencite{Piepho2013Why}, however this is not true when the goal of OFE shifts from what is desired from a conventional analysis. In the application of precision agriculture (PA), a prescription map from the experimental results is required by the variable-rate applicators (VRA) prior to the start of the operation \parencite{Pringle2004FieldScale}. Therefore, in this scenario, the goal of OFE becomes obtaining such a smooth map showing the optimal level of a controllable input, such as nitrogen rates, across a grid made by rows and columns covering the whole field. For this objective, \textcite{Pringle2004FieldScale} stress that only a single level of the treatments can be directly observed at any one point on the grid, and the response for other levels at the same grid must be interpolated. If a randomised design is conducted, the interpolation distances to locations with treatment levels of interest will vary across the field. Such heterogeneous distances increase the uncertainty in the analysis and reduce the efficiency of local prediction \parencite{Piepho2011Statistical}. Hence, for this scenario a systematic design is preferable to a randomised design. Unfortunately, this standpoint has been ignored by researchers, meaning that randomised designs have been universal.

To analyse a systematic design for the creation of the optimal treatment map for OFE is statistically challenging. The truly localised estimation at each point on the grid is unknown and the optimum treatment response continuously varies spatially on the field. \textcite{Cao2022Bayesian} implemented a Bayesian approach of spatially correlated random parameters for large systematic OFE strip trials. They assumed a quadratic response in the model, consisting of a global component and a local spatially varying component. However, the Bayesian approach is time-consuming and requires preliminary knowledge of Bayesian statistics for farmers and agronomists. Alternatively, \textcite{Rakshit2020Novel} adapted a local regression approach, called geographically weighted regression (GWR), to obtain spatially-varying estimates of treatment effects for OFE. Additionally, \textcite{Evans2020Assessment} conclude that, through simulation studies, GWR is a simple method for the analysis of OFE data and is able to accurately separate yield variation that is not due to the applied treatment from yield response due to treatment. The limitation in the study is that they used a randomised design and assumed a linear response to fertiliser treatment. 

%The role of variable-rate applicators has a focus on the application of fertilizers or fungicides to target enable the most efficient yields. 

\textcite{Piepho2018Tutorial} demonstrate an example that a linear model is lack of fit on the sugar beets data \parencite{Petersen1994Agricultural}. \textcite{Glynn2007} show that many curves exist beyond a linear trend for nutrient-response relationships. 
%However, when modelling this process it is still common to see linear approximations made \parencite{}. 
%While segments of nutrient-response relationships may be linear, they will generally exhibit inflection points which arise from processes such as sufficiency or toxicity. 
The curve depends on the current availability of other macro and micro nutrients in the soil \parencite{Marschner2011}, meaning that a linear relationship is unlikely to be consistent across a large trial. For this reason, it is important to consider models with degrees higher than 1, with a quadratic model found to be adequate relationships (\textcite{Piepho2018Tutorial, Liben2019}). 

%Other relationships have been  with the potential for other more complex polynomials possible but can be prone to over-fitting. %\ref{Add a reference}. 

% nature than linear curves. 
% 

Our study uses a few simulation examples to demonstrate that randomisation is not a crucial prerequisite for large strip trials, and, in the purpose of obtaining a treatment map, a systematic design is superior over a randomised design subject to appropriate restrictions. We also test the power of GWR, allowing us to know if it can successfully estimate spatially varying treatment effects for both linear and quadratic responses to treatments. It shows that the optimal bandwidths found by AICc is not the best bandwidth for GWR, but a fixed bandwidth based on the experimental design is recommended.

The structure of the paper is organised as follows: in Section \ref{Sec:Meth}, we describe the statistical model for generating simulated data, which has spatially varying coefficients of treatments, and the GWR model for fitting OFE data. In Section \ref{Sec:Simu}, we generate simulated data for the combination of the following scenarios: randomised and systematic designs; linear and quadratic response; low and high coefficient correlations; spatial variation among grids is identity (no spatial trend), $\AR\otimes\AR$ and \Matern form. Finally, in Sections \ref{Sec:Res} and \ref{Sec:Dis}, we illustrate the results and discuss their importance with respect to OFE, and how the findings should influence future trial designs.

\section{Methods}\label{Sec:Meth}

This section describes the statistical model used in the simulation study. It outlines the basic model (subsection \ref{sec:basic}) followed by the methodology for the spatially correlated treatment parameters (subsection \ref{sec:spatial}), and finally GWR (subsection \ref{sec:gwr}).

%and explains the statistical aspects of the models with spatially correlated coefficients.

\subsection{Basic statistical model}\label{sec:basic}

In a conventional agricultural study, a field experiment can be considered as a rectangular array, consisting of $r$ rows and $c$ columns, where the total number of the plots in the experiment is $n=r\times c$. The notation $s_i\in \mathcal{R}^2, i=1,\ldots,n$ is a two-cell vector of the Cartesian coordination of the plot centroids, located on a regular grid \parencite{Zimmerman1991Randoma}. Hence, $y(s_i)$ denotes the dependent variable at a query location/grid $i$.

With the assumption that $\bm{Y}$ is the vector of the plot data ordered as rows nested within ranges (columns), then the matrix notation of the model is 
\begin{equation}\label{eq:modelmatrix}
	\bm{Y} = \bm{X}\bm{b}+\bm{Z}\bm{u}+\bm{e},
\end{equation}
where $\bm{b}$ and $\bm{u}$ are vectors of fixed and random effects, respectively; $\bm{X}$ and $\bm{Z}$ are the associated design matrices; and $\bm{e}$ is the error vector. We further assume that $\bm{u}$ and $\bm{e}$ are pairwise independent and that their joint distribution is 
\begin{equation}\label{eq:covariance}
	\begin{bmatrix}
		\bm{u} \\ \bm{e}
	\end{bmatrix} \sim \N\left( \begin{bmatrix}
		0\\0 \end{bmatrix}, \begin{bmatrix}
		\Sigma_u & 0 \\ 0 & \Sigma_e
	\end{bmatrix}\right). 
\end{equation}

\subsection{Spatially correlated treatment parameters}\label{sec:spatial}

\textcite{Cao2022Bayesian} implemented the Bayesian hierarchical model with spatially correlated random parameters on large OFE strip trials. Here, for the simulation study, we use the same model to generate the simulated data. 
	
With the same notation given in the reference and previous section, the underlying model is represented as 
\begin{equation}\label{eq:underlying}
	\begin{split}
		y(s_i)\mid \bm{u}_i,\theta_u,\sigma_e &\sim \N\left( \sum_{m=1}^{l}b_m x_m(s_i) + \sum_{j=1}^{k}u_j(s_i)z_j(s_i),~e(s_i)\right) \\
		\bm{u}_i \mid \theta_u &\sim \N(0,V_u(\theta_u))\\
		e(s_i) \mid \sigma_e &\sim \N(0,\sigma_e^2) 
	\end{split}
\end{equation}
where: $x_1,\ldots, x_l$ denote $l$ fixed effects and $z_1,\ldots, z_k$ denote $k$ random effects; $b_m$ and $u_j(s)$ are the coefficients for the fixed and random terms, respectively; $\bm{u}_i$ is a vector of all random effects at grid $s_i\in\mathcal{S}$, $i=1,\ldots,n$; $\theta_u$ is a set of parameters of the covariance matrix $V_u$; and $\sigma_e$ is a positive latent variable.

In \eqref{eq:underlying}, the structure of the covariance matrix $V_u(\theta_u)$ of $\bm{u}_i$ can be either diagonal, which means the treatments at grid $i$ are independent, or in general form, which means a correlation exists. \textcite{McElreath2015Statistical} suggest that the covariance of $\bm{u}_i$ can be $V_u = B(\sigma_u)R_uB(\sigma_u)$, where $B(\sigma_u)$ denotes the diagonal matrix of elements $\sigma_{u_j}$, $j=1,\ldots,k$, and $R_u \sim \mbox{LKJ}(\epsilon)$ is a correlation matrix controlled by a positive parameter $\epsilon$. As $\epsilon$ increases, a high correlation among parameters becomes less likely.

Furthermore, by incorporating a spatial correlation structure $V_s$, the complete form of the covariance matrix of $\bm{u}$ is presented as 
\begin{equation}\label{eq:varu}
	\Sigma_u = V_s \otimes V_u. 
\end{equation}
In fact, $V_s$ is the covariance matrix of all grids on the field. For example, if $V_s=I_{n\times n}$ is an identity matrix, each grid is independent even though the treatments within each grid are correlated. However, the correlation among grids is ubiquitous. Hence, we introduce a simple spatial covariance matrix such that 
\begin{equation}\label{eq:ar1cov}
	V_s = \AR(\rho_c)\otimes \AR(\rho_r),
\end{equation}
where AR1 is the separable first-order autoregressive model in the column and row direction which is controlled by the correlation parameters $\rho_c$ and $\rho_r$, respectively \parencite{Butler2017ASRemlR}. 

Besides the above $\AR\otimes\AR$ covariance, the \Matern class covariance  
\begin{equation}\label{eq:matcov}
V_s(d) = \sigma^2 \frac{2^{1-\nu}}{\Gamma(\nu)} \left( \sqrt{2\nu} \frac{d}{r}\right)^\nu K_\nu\left( \sqrt{2\nu} \frac{d}{r}\right)
\end{equation}
is also used in spatial analysis \parencite{Cressie1999Classes} and in capturing spatial variation in OFE \parencite{Selle2019Flexible}. Here, $d$ is the space lag or distance; $r$ is a non-negative scaling parameter; $\nu> 0$ is a smoothness parameter determining the mean-square differentiability of the field; $\sigma_d^2$ is the variance of the process; $\Gamma$ is the Gamma function; and $K_\nu$ is the modified Bessel function of the second kind. If $\nu = r + \frac{1}{2}$, then the \Matern covariance can be expressed as a product of an exponential and a polynomial of order $r$ \parencite{Pandit2019Comparative, Abramowitz1974Handbook}, which simplifies the model and the computation process. The \Matern class $\nu=\frac{3}{2}$ and $\nu =\frac{5}{2}$ are common in application.

Model \eqref{eq:underlying} has the advantage of reproducibility in simulation study and robustness in estimation. It is possible even though only a single treatment is directly observed at each plot, and the responses of the other levels need to be estimated by interpolation because the spatial model allows using information from neighbouring plots with other treatments \parencite{Panten2010Enhancing, Piepho2011Statistical}.

\subsection{Geographically weighted regression (GWR)}\label{sec:gwr}

Geographically weighted regression (GWR) is a local regression approach and is adapted to obtain spatially-varying estimates of treatment effects for OFE \parencite{Rakshit2020Novel}.  It is seen as a locally weighted regression method that operates by assigning a weight to each observation $i$ depending on its distance from the query grid on the field \parencite{Paez2002General}.

The underlying template model for the GWR, according to \textcite{Leung2000Statistical}, is given by 
\begin{equation}\label{eq:basicgwr}
	y(s_i) = \beta_0 + \sum_{j=1}^{k} \beta_jz_j(s_i)+\varepsilon_i, 
\end{equation}
where $\bm{\beta}$ and $\varepsilon\sim \N(0,\tau^2)$ are the model parameters for the $k$ levels treatments and error terms, respectively, at grid $i,\ i=1,\ldots,n$. The estimator of this model is given by the geographically weighted expression in 
\begin{eqnarray}\label{eq:betahat}
	\bm{\hat{\beta}}(s) = \left( Z^\top W(s) Z \right)^{-1}Z^\top W(s) Y,
\end{eqnarray}
where $W(s)$ is an $n\times n$ diagonal matrix of weights, $\bm{w}$. Then it can be found by maximising the local log-likelihood 
\begin{eqnarray}\label{eq:locallog}
	\log L(s;\beta) = - \frac{1}{2\tau^2}\sum_{i=1}^{n}K(s,s_i)\left( y(s_i) -\beta_0 - \sum_{j=1}^{k}\beta_jz_j(s_i) \right)^2 
\end{eqnarray}
with a given kernel function $K(\cdot,\cdot)$, such as Gaussian, exponential, bi-square or tri-cube \parencite{Gollini2015GWmodel}. In the simulation study, we use Gaussian kernel. In fact, kernel is not a crucial factor in GWR model fitting on OFE data. By contrast, the factor bandwidth has higher influence on the estimation.

The optimal bandwidth of a GWR model is usually given by the lowest AICc where 
\begin{equation}\label{eq:aicc}
	\mbox{AICc} = 2n\log (\tau^2) + n \log (2\pi) + n\frac{ n+\tr (S) }{n-2-\tr (S)},
\end{equation}
and $S$ is the matrix with the $i$-th row given by $Z_i\left( Z^\top W(s_i) Z \right)^{-1}Z^\top W(s_i)$ \parencite{Evans2020Assessment}. Alternatively, as suggested by \textcite{Rakshit2020Novel}, it can be based on the experimental design such that the local regressions capture data covering the full range of treatments.

The GWR model in this paper is implemented with the \R-package \texttt{GWmodel} \parencite{lu2014gwmodel, Gollini2015GWmodel}.

\section{Simulation study}\label{Sec:Simu}

To study the effect of randomised designs and systematic designs and to evaluate the power of GWR, we simulate spatially correlated large strip trials. The advantage of the simulation study is that the actual coefficients of the models are known, so adverse effects of model misspecification
can be ruled out \parencite{Piepho2013Why}. 

We investigate the combination of the following factors: types of designs with two levels: randomised and systematic; response relationship with two levels: linear and quadratic; correlation of coefficients with two levels: low and high; spatial variation with three levels: identity (no spatial trend), $\AR\otimes\AR$ and \Matern form; bandwidth of GWR with three levels: 5, 9 and optimum given by AICc. The fixed bandwidth 5 in the simulation study covers all treatment levels (five nitrogen levels) in a systematic design, where the information is adequate in the inference of a quadratic curve. Similarly, the fixed bandwidth 9 covers all possible treatment levels in a randomised design. This is because if all treatments are randomly allocated in the strips, there is a chance that the information of one treatment level is missing if the bandwidth is less than 9. 

\begin{figure}[!htp]
	\begin{subfigure}[t]{0.45\textwidth}
		\centering
		\includegraphics[width=\linewidth]{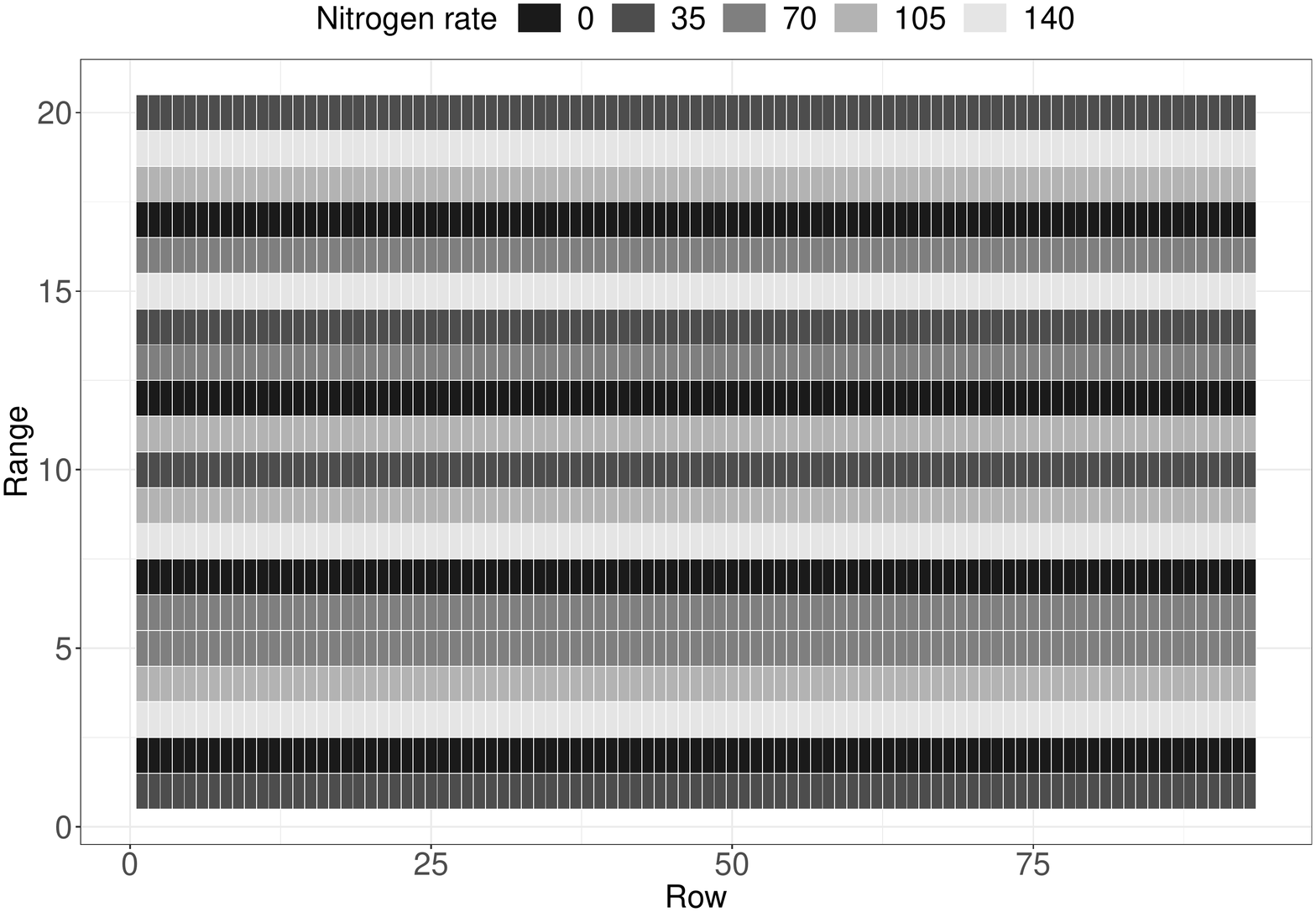}
		\caption{Treatments are randomly allocated into large strips.}
	\end{subfigure}
	\hspace{0.05\textwidth}
	\begin{subfigure}[t]{0.45\textwidth}
		\centering
		\includegraphics[width=\linewidth]{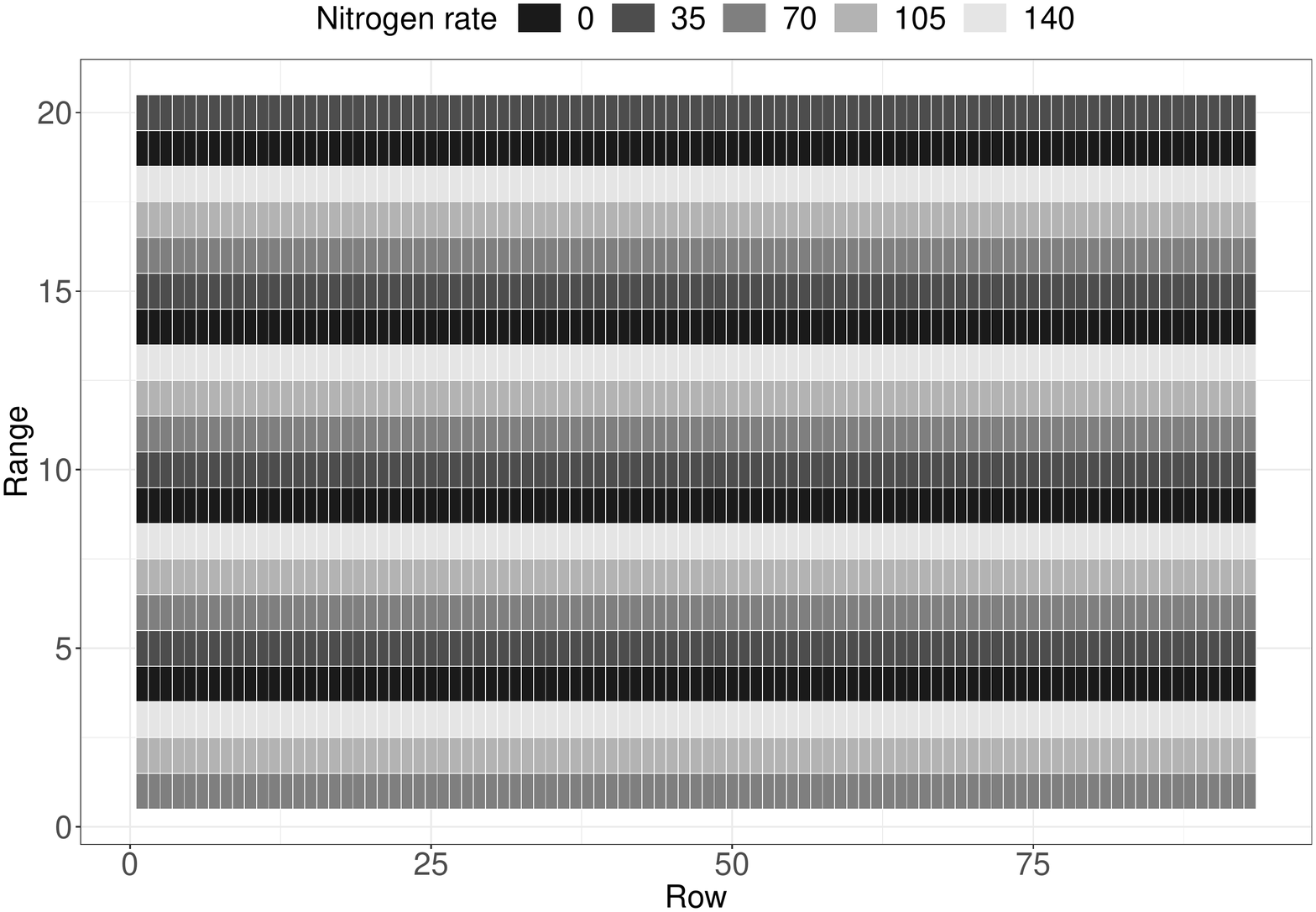}
		\caption{Treatments are systematically allocated into large strips.}
	\end{subfigure}
	\caption{The nitrogen treatments with five levels (0, 35, 70, 105 and 140 kg/ha) randomly (a) and systematically (b) allocated into strips.}\label{fig:Nitrogen}
\end{figure}

With model \eqref{eq:underlying}, we generate the yield as a response to the nitrogen treatment in two scenarios which cover whether the yield has a linear or quadratic relationship with nitrogen rate. The nitrogen rates are treated as continuous observations with five levels: 0, 35, 75, 105 and 140 kg/ha. A strip plot structure was used to allocate the five nitrogen levels where each level is assigned to one strip. Then we assume that the experimental design of the trial consists of four replicates, each containing 5 ranges (columns) by 93 rows. The nitrogen rates are allocated either randomly or systematically on the whole field. The overall layout of the trial is 20 ranges by 93 rows. Examples of a randomly and systematically allocated treatment map are presented in Figure \ref{fig:Nitrogen}.

With model \eqref{eq:underlying}, the nitrogen rates are treated as continuous observations with five levels. For a linear relationship, we assume that the true global intercept is $b_0 = 65$ and the slope is $b_1=0.05$. The parameters are chosen according to the estimates by \textcite{Rakshit2020Novel} on the Las Rosas corn field data from the \texttt{R}-package \texttt{agridat} \parencite{White2008Agridat}. The variances of $\bm{u}_i$ are $\sigma_{u_0} = 5$ and $\sigma_{u_1}=0.01$. For the $\AR\otimes\AR$ covariance matrix in \eqref{eq:ar1cov}, the two correlation parameters for column and row are $\rho_c = 0.15$ and $\rho_r=0.5$. We have assumed a higher correlation in the rows due to the fact that the crop are traditionally sown and harvested along ranges where the correlation is higher in perpendicular to the travel direction \parencite{Marchant2019Establishinga}. For the \Matern covariance matrix \eqref{eq:matcov}, we set $\sigma_d^2=1$, $r=1$ and $\nu = \frac{3}{2}$. After drawing samples of $\bm{u}$ from $\N(0,\Sigma_u)$, the true spatially varying coefficients are $\bm{\beta}_0 = b_0 + \bm{u}_0$ and $\bm{\beta}_1 = b_1 + \bm{u}_1$. 

Similarly, for the quadratic relationship, we have the true global intercept $b_0 = 65$, coefficients $b_1 = 0.05$ and $b_2 = -0.0003$, 
%where the curve is in a frown shape. 
making the curve concave down. We keep $\sigma_{u_0} = 5$, $\sigma_{u_1}=0.01$ and add $\sigma_{u_2}=0.0001$ because the true values are small. The remaining parameters stay unchanged. Therefore, the true spatially varying coefficients are $\bm{\beta}_0 = b_0 + \bm{u}_0$, $\bm{\beta}_1 = b_1 + \bm{u}_1$ and $\bm{\beta}_2 = b_2 + \bm{u}_2$.

\begin{figure}[H]
	\begin{subfigure}[t]{0.45\textwidth}
		\centering
		\includegraphics[width=\linewidth]{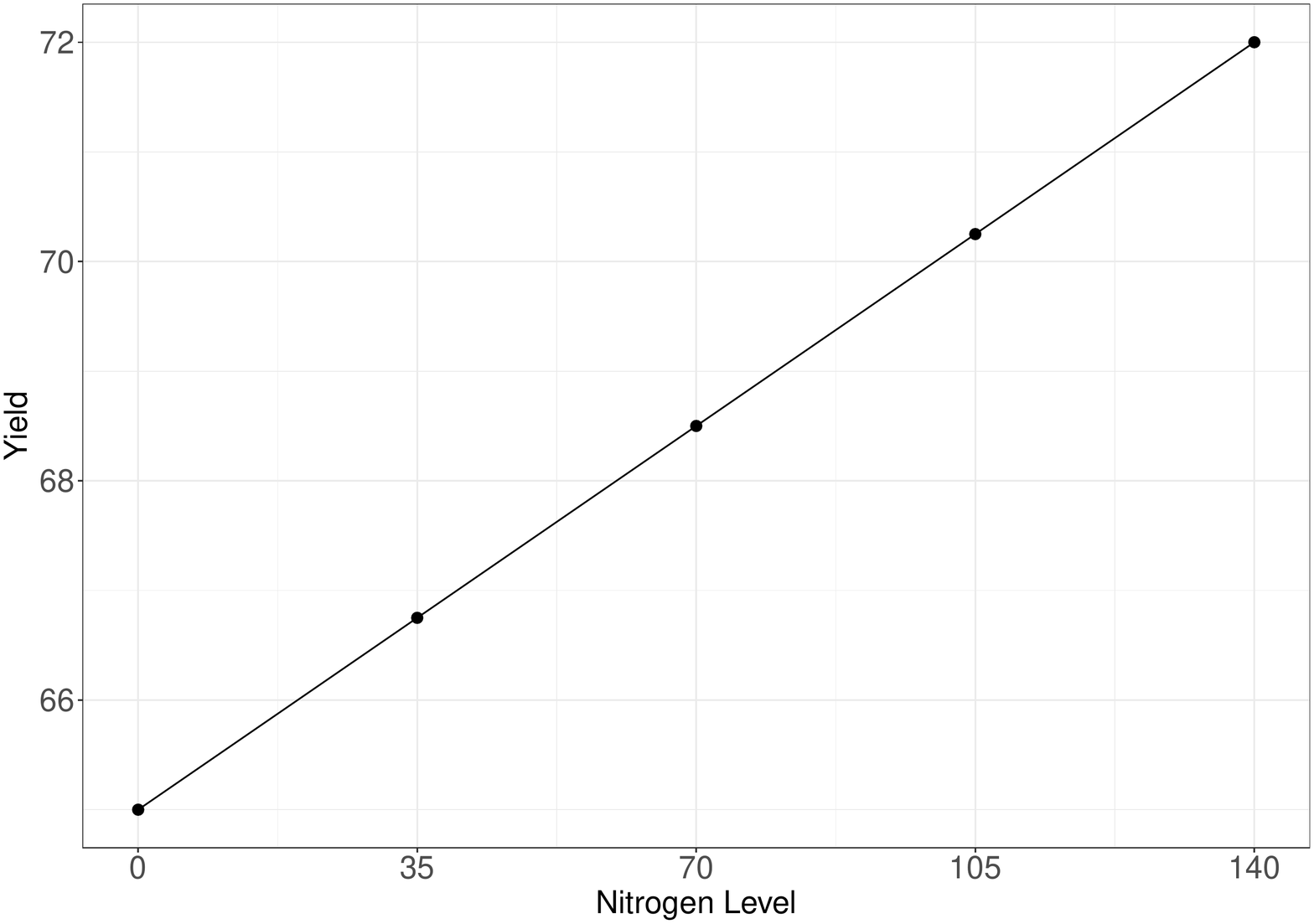}
		\caption{Linear relationship of yield to nitrogen level.}
	\end{subfigure}
	\hspace{0.05\textwidth}
	\begin{subfigure}[t]{0.45\textwidth}
		\centering
		\includegraphics[width=\linewidth]{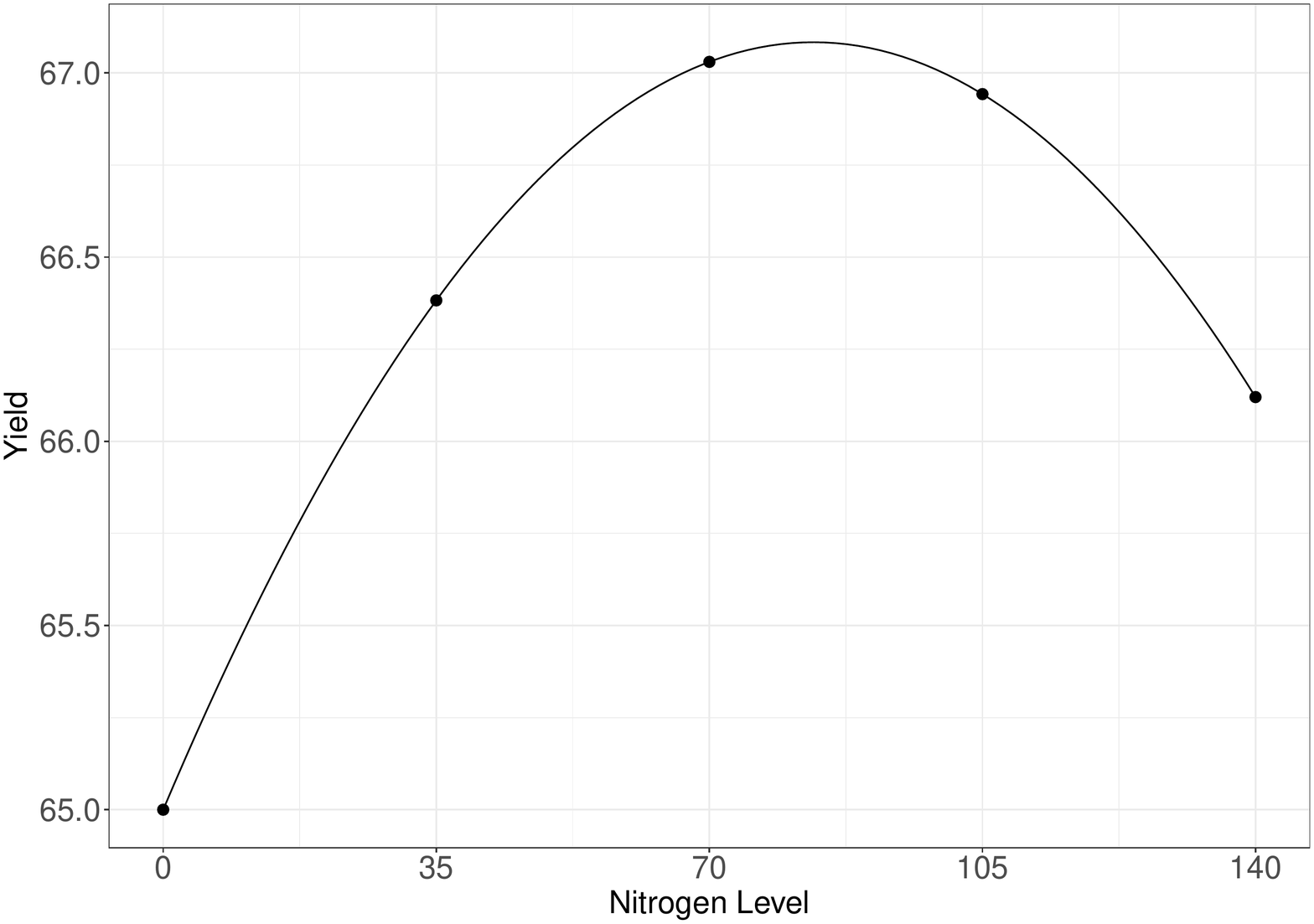}
		\caption{Quadratic relationship of yield to nitrogen level.}
	\end{subfigure}
	\caption{Noise-free linear and quadratic response to fertiliser treatment.}\label{fig:Lines}
\end{figure}

To summarise, with the true spatially varying coefficients of the treatments, the simulated yield is obtained by 
\begin{equation}
\begin{cases}
	% \text{Linear}  &y_i = b_0 + b_1N_i + u_{0i} + u_{1i}N_i + e_i \\
	\text{Linear}  &y_i = b_0 + u_{0i} + (b_1 + u_{1i})N_i + e_i \\
	% \text{Quadratic} &y_i = b_0 + b_1N_i + b_2N_i^2 + u_{0i} + u_{1i}N_i  + u_{2i}N_i^2 + e_i
	\text{Quadratic} &y_i = b_0 + u_{0i} + (b_1 + u_{1i})N_i + (b_2 + u_{2i})N_i^2 + e_i
\end{cases},
\end{equation}
where $N_i$ is the nitrogen rate, $e_i\sim \N(0,1)$ is the error term at grid $i$, $i = 1, \ldots, n$. Figure \ref{fig:Lines} illustrates how these curves behave for the linear and quadratic relationships.

% Examples of the yield maps with true $\AR\otimes\AR$ spatially varying coefficients are in Figures \ref{fig:LinearYieldMap} and \ref{fig:QuadYieldMap}. 
% \begin{figure}[!htp]
% 	\centering
% 	\begin{subfigure}[t]{0.45\textwidth}
% 		\centering
% 		\includegraphics[width=\linewidth]{Expt/LinearRandYield01.pdf}
% 		\caption{Yield map of the linear response of the randomised design. }
% 	\end{subfigure}
% 	\begin{subfigure}[t]{0.45\textwidth}
% 		\centering
% 		\includegraphics[width=\linewidth]{Expt/LinearSystYield01.pdf}
% 		\caption{Yield map of the linear response of the systematic design. }
% 	\end{subfigure}
% 	\caption{Yield maps of linear response of randomly and systematically allocated nitrogen rates.}\label{fig:LinearYieldMap}
% \end{figure}

% \begin{figure}[!htp]
% 	\centering
% 	\begin{subfigure}[t]{0.45\textwidth}
% 		\centering
% 		\includegraphics[width=\linewidth]{Expt/QuadRandYield01.pdf}
% 		\caption{Yield map of the quadratic response of the randomised design. }
% 	\end{subfigure}
% 	\begin{subfigure}[t]{0.45\textwidth}
% 		\centering
% 		\includegraphics[width=\linewidth]{Expt/QuadSystYield01.pdf}
% 		\caption{Yield map of the quadratic response of the systematic design. }
% 	\end{subfigure}
% 	\caption{Yield maps of quadratic response of randomly and systematically allocated nitrogen rates.}\label{fig:QuadYieldMap}
% \end{figure}

%% To fit the data, we use basic GWR with Gaussian kernels. 

The purpose of the simulation study is to test the effect of different types of designs in coefficients estimation by GWR. Identical coefficients were used in one comparison process for two types of designs, and the yield reflects the effect of nitrogen rates.

\section{Results}\label{Sec:Res}

By running the simulation 100 times, we assessed the performance of the randomised and systematic designs for linear and quadratic responses. In subsection \ref{Sec:MSE} the mean squared errors are compared between the two designs for different bandwidths, parameter correlations and spatial covariance matrices. Subsection \ref{Sec:anova} uses an analysis of variance (ANOVA) test to explore the significance of the factors in the simulation, while subsection \ref{Sec:bandselect} states the performance of bandwidth selection using AICc. 

%By running the simulation 100 times, 

\subsection{Mean squared error}\label{Sec:MSE}

\begin{figure}[H]
	\centering
	\includegraphics[width=\linewidth]{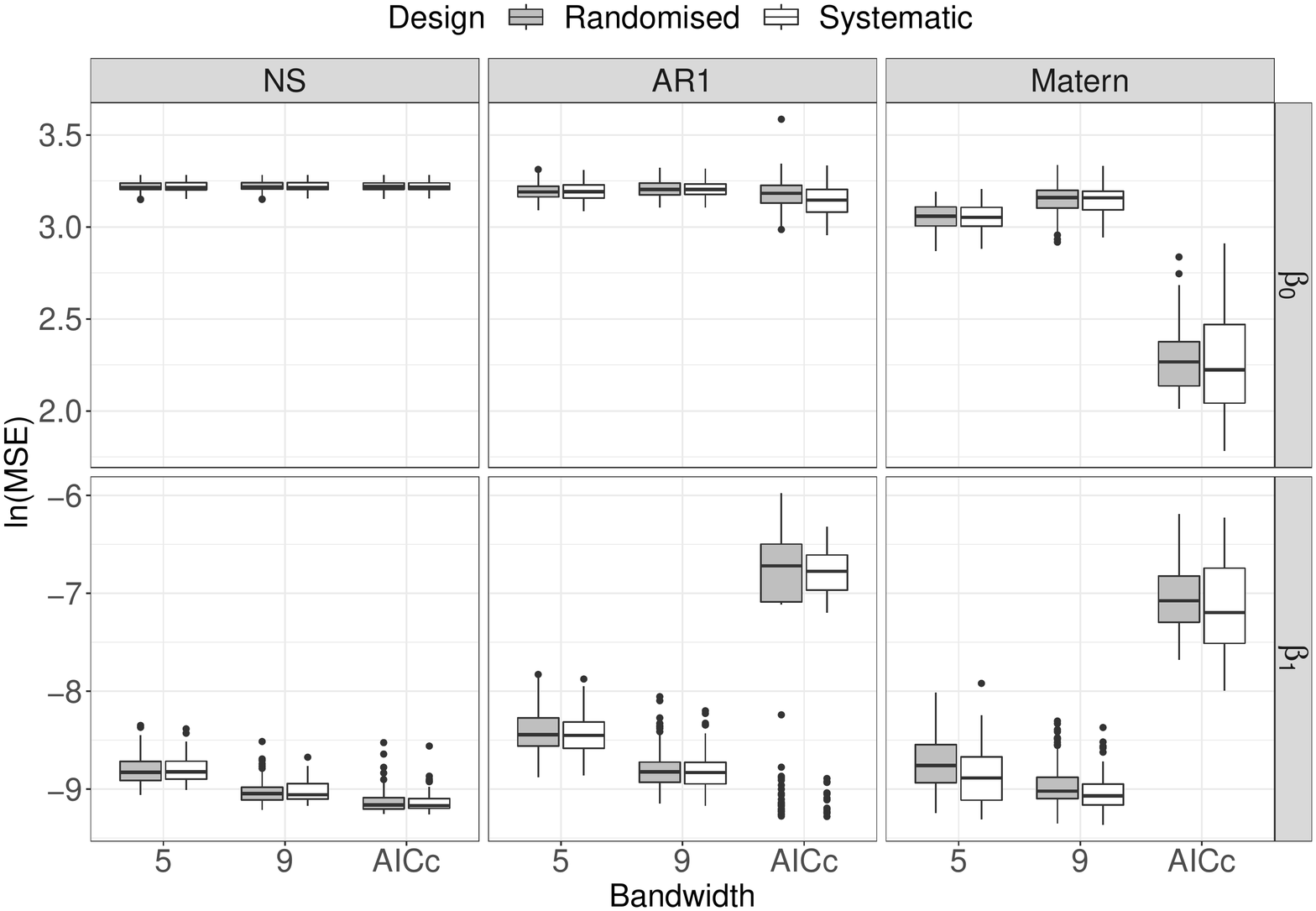}
	\caption{Boxplots of $\ln$(MSE) for $\hat{\beta}_0$ and $\hat{\beta}_1$ in GWR models using different bandwidths for the simulated data with a linear response. The simulated data had different spatial covariance matrices (NS, AR1$\otimes$AR1 and \Matern) and a low correlation between the parameters ($\epsilon=1$).}\label{fig:LinBetaMSE}
\end{figure}

\begin{figure}[H]
	\centering
	\includegraphics[width=\linewidth]{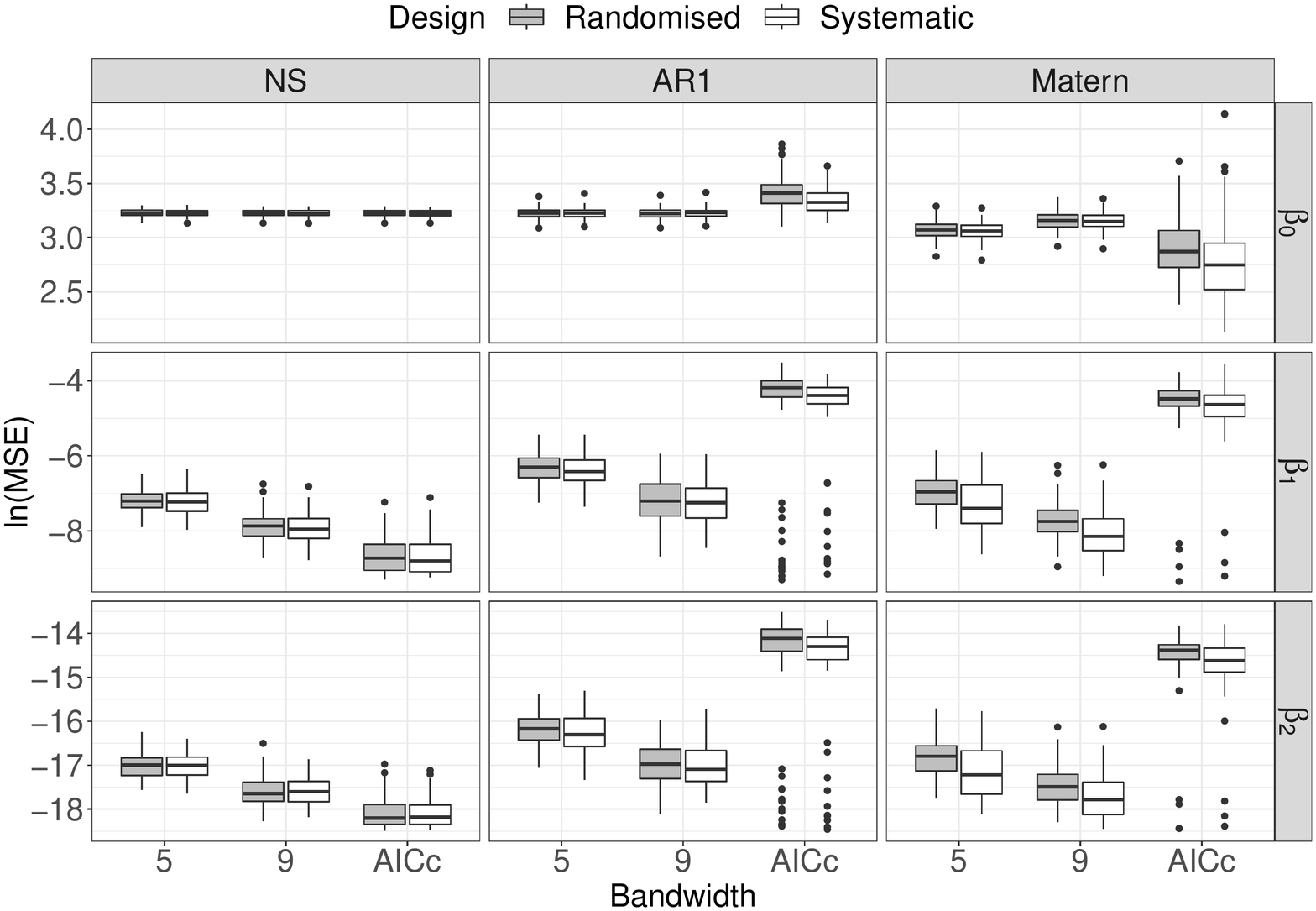}
	\caption{Boxplots of $\ln$(MSE) for $\hat{\beta}_0$, $\hat{\beta}_1$ and $\hat{\beta}_2$ in GWR models using different bandwidths for the simulated data with a quadratic response. The simulated data had different spatial covariance matrices (NS, AR1$\otimes$AR1 and \Matern) and a low correlation amongst the parameters ($\epsilon=1$).} \label{fig:QuadBetaMSE}
\end{figure}

We assessed the true mean squared error (MSE) of estimated coefficients differences. This was calculated by the difference of true coefficients, $\bm{\beta} = b + \bm{u}$, and estimated spatially varying coefficients, $\bm{\hat{\beta}} = \hat{b}+\bm{\hat{u}}$, for each grid, and then squared the discrepancy and averaged across the field for comparison. The results are shown in Figures \ref{fig:LinBetaMSE} and \ref{fig:QuadBetaMSE} where ``NS'' stands for no spatial variation ($V_s=I_{n\times n}$), ``AR1'' is for $\AR(0.15)\otimes \AR(0.5)$ and ``Matern'' is the \Matern covariance with $\nu=\frac{3}{2}$. Since the MSE of $\beta_0$, $\beta_1$ and $\beta_2$ are small, we take the natural logarithm for better visualisation. 

% Paragraph about Figure 3 and the performance for a linear response
With the assumption of a linear response, both randomised and systematic designs performed similarly, meaning GWR is able to partition the local varying intercept and treatment coefficient. Figure \ref{fig:LinBetaMSE} shows that if the response is linear, the MSE of $\hat{\beta}_0$ and $\hat{\beta}_1$ estimated by GWR
%with bandwidth 5, 9 and the optimal bandwidth found by AICc
, for all bandwidths,
is not distinguishing between randomised and systematic designs. This is true regardless of the type of spatial covariance matrix when the correlation within grids is small ($\epsilon=1$), or high ($\epsilon=0.1$) as is shown in Figure \ref{fig:LinBetaMSEeta01}. The GWR with bandwidth selected by AICc had a smaller MSE for its coefficients than the model with a fixed bandwidth (Figure \ref{fig:LinBetaMSE}).

% Paragraph about Figure 4 and the performance for a linear response
However, if the assumption is a quadratic response, Figure \ref{fig:QuadBetaMSE} and \ref{fig:QuadBetaMSEeta01} show that the GWR with a fixed bandwidth of a systematic design outperforms a randomised design if the spatial correlation is taken into account. With the optimal bandwidth, GWR successfully estimates the global intercepts $\beta_0$, but failed in estimating local varying coefficients $\beta_1$ and $\beta_2$, where the MSE is relatively larger than with fixed bandwidth. However, if we only compare across two types of designs, it still proves that GWR is robust to fit a systematic design rather than a randomised design if the assumption is quadratic response, regardless the intensity of the correlation within grids. 

% Bandwidth MSE Result

The relative MSE for each bandwidth was found to change according to the spatial covariance matrix as well as the coefficient. When no spatial variation was simulated for $\beta_1$ and $\beta_2$, the bandwidth selected by AICc had the lowest MSE, followed by the bandwidths 9 and then 5. When there was spatial variation (either AR1$\otimes$AR1 or \Matern) present the estimation for $\beta_1$ and $\beta_2$ had lowest MSE when using bandwidth 9, followed by 5 and then bandwidth from AICc. The intercept ($\beta_0$) showed only significant changes in MSE amongst the bandwidths when the \Matern spatial covariance was considered. In this case, the bandwidth selected by AICc produced the smallest MSE.

Tables \ref{tb:MSElinear} and \ref{tb:MSElinearHigh} are the median MSE of linear response for two scenarios: correlation is low and correlation is high.

\begin{table}[H]
	\centering
\begin{threeparttable}
	\caption{Median MSE of GWR coefficient estimates for a linear response when the correlation between the parameters is low ($\epsilon=1$).}\label{tb:MSElinear}
	\begin{tabular}{llcccccc}
		\toprule
		&  & \multicolumn{3}{c}{Randomised} & \multicolumn{3}{c}{Systematic} \\ 
   		 & Coefficient & 5  &  9  & AICc & 5   & 9  & AICc \\ \midrule
		\multirow{2}{*}{NS}   & $\hat{\beta}_0$ & 24.903 &	24.924 &	24.983&	24.886\tnote{$\dagger$} &	24.911 &	24.965 \\ 
		& $\hat{\beta}_1 (\times 10^3)$ & 0.147&	0.118&	0.105&	0.147&	0.117&	0.104\tnote{$\dagger$}  \\  \midrule
		\multirow{2}{*}{AR1}  & $\hat{\beta}_0$ & 24.308&	24.617&	24.126&	24.319&	24.617&	23.246\tnote{$\dagger$}  \\ 
		& $\hat{\beta}_1 (\times 10^3)$ & 0.215	&0.147	&1.208&	0.214&	0.146\tnote{$\dagger$}	&1.143 \\ \midrule
		\multirow{2}{*}{\Matern} & $\hat{\beta}_0$ & 21.303	&23.566	&9.647&	21.164&	23.526&	9.239\tnote{$\dagger$} \\ 
		& $\hat{\beta}_1 (\times 10^3)$ & 0.157	&0.121&	0.845&	0.138&	0.115\tnote{$\dagger$} &	0.749 \\
		\bottomrule
	\end{tabular}
	\begin{tablenotes}
	\item[$\dagger$] \footnotesize Indicates the smallest MSE for the row.
	\end{tablenotes}
\end{threeparttable}
\end{table}

\begin{table}[H]
	\centering
\begin{threeparttable}
	\caption{Median MSE of GWR coefficient estimates of linear response when the correlation between the parameters is high ($\epsilon=0.1$).}\label{tb:MSElinearHigh}
	\begin{tabular}{llcccccc}
		\toprule
		&  & \multicolumn{3}{c}{Randomised} & \multicolumn{3}{c}{Systematic} \\ 
		& Coefficient & 5  &  9  & AICc & 5   & 9  & AICc \\ \midrule
		\multirow{2}{*}{NS}   & $\hat{\beta}_0$  & 24.961	&24.991	&24.958	&24.934\tnote{$\dagger$}	&24.992	&25.020 \\
		& $\hat{\beta}_1 (\times 10^3)$ & 0.144&	0.115&	0.103\tnote{$\dagger$} &	0.145&	0.116&	0.104 \\ \midrule
		\multirow{2}{*}{AR1$\otimes$AR1}  & $\hat{\beta}_0$  & 24.234&	24.518&	23.696&	24.171	&24.497&	23.421\tnote{$\dagger$}  \\
		& $\hat{\beta}_1 (\times 10^3)$   & 0.216	&0.144\tnote{$\dagger$} &	1.024&	0.211&	0.146&	1.027 \\ \midrule
		\multirow{2}{*}{\Matern} & $\hat{\beta}_0$  & 20.882&	22.811&	9.823&	20.770	&22.789	&9.257\tnote{$\dagger$}  \\
		& $\hat{\beta}_1 (\times 10^3)$  & 0.152&	0.120&	0.939&	0.140&	0.112\tnote{$\dagger$} &	0.815 \\ 
		\bottomrule
	\end{tabular}
		\begin{tablenotes}
	\item[$\dagger$] \footnotesize Indicates the smallest MSE for the row.
	\end{tablenotes}
\end{threeparttable}
\end{table}

\vspace{12pt}

Tables \ref{tb:MSEquadratic} and \ref{tb:MSEquadraticHigh} are the median MSE of quadratic response for two scenarios: correlation is low and correlation is high. Despite the correlation intensity, the data fitted by GWR from systematical designs are superior to randomised designs, having a smaller MSE.

\begin{table}[H]
	\centering
\begin{threeparttable}
	\caption{Median MSE of GWR coefficient estimates of quadratic response when the correlation amongst the parameters is low ($\epsilon=1$).}\label{tb:MSEquadratic}
	\begin{tabular}{llcccccc} \toprule
		 &  & \multicolumn{3}{c}{Randomised} & \multicolumn{3}{c}{Systematic} \\ 
		  & Coefficient & 5  & 9  & AICc & 5  & 9  & AICc \\ \midrule
		\multirow{3}{*}{NS}   & $\hat{\beta}_0$ & 25.218 & 25.150 & 25.152  & 25.263 & 25.179 & 25.138\tnote{$\dagger$}  \\
		& $\hat{\beta}_1 (\times 10^4)$ & 7.417  & 3.823  & 1.625 & 7.233  & 3.529  & 1.516\tnote{$\dagger$}  \\
		& $\hat{\beta}_2 (\times 10^8)$  & 4.157  & 2.168  & 1.242\tnote{$\dagger$} & 4.135  & 2.269  & 1.269   \\ \midrule 
		\multirow{3}{*}{AR1$\otimes$AR1}  & $\hat{\beta}_0$  & 25.185 & 25.092\tnote{$\dagger$} & 30.315  & 25.166 & 25.230 & 27.831  \\
		&$\hat{\beta}_1 (\times 10^4)$  & 18.395 & 7.414  & 151.595 & 16.243 & 7.124\tnote{$\dagger$}  & 123.181  \\
		& $\hat{\beta}_2 (\times 10^8)$ & 9.491  & 4.244  & 74.420  & 8.305  & 3.777\tnote{$\dagger$}  & 61.619 \\ \midrule
		\multirow{3}{*}{\Matern} & $\hat{\beta}_0$  & 21.532 & 23.502 & 17.680  & 21.384 & 23.319 & 15.631\tnote{$\dagger$} \\
		&$\hat{\beta}_1 (\times 10^4)$  & 9.502  & 4.326  & 112.914 & 6.121  & 2.901\tnote{$\dagger$}  & 96.829  \\
		& $\hat{\beta}_2 (\times 10^8)$  & 5.071  & 2.537  & 56.789  & 3.324  & 1.889\tnote{$\dagger$}  & 44.707 \\ \bottomrule
	\end{tabular}
	    \begin{tablenotes}
	        \item[$\dagger$] \footnotesize Indicates the smallest MSE for the row.
	    \end{tablenotes}
\end{threeparttable}
\end{table}

\begin{table}[H]
	\centering
\begin{threeparttable}
	\caption{Median MSE of GWR coefficient estimates of quadratic response when the correlation amongst the parameters is high ($\epsilon=0.1$).}\label{tb:MSEquadraticHigh}
	\begin{tabular}{llcccccc} \toprule
		&  & \multicolumn{3}{c}{Randomised} & \multicolumn{3}{c}{Systematic} \\ 
		$\epsilon=0.1$  & Coefficients & 5  & 9  & AICc & 5  & 9  & AICc \\ \midrule
		\multirow{3}{*}{NS} & $\hat{\beta}_0$  & 25.075 & 25.067 & 25.015  & 25.082 & 25.060 & 25.012\tnote{$\dagger$} \\
		&$\hat{\beta}_1 (\times 10^4)$  & 6.683  & 3.466  & 1.478\tnote{$\dagger$} & 7.353  & 3.472  & 1.506 \\
		&$\hat{\beta}_2 (\times 10^8)$  & 3.779  & 2.101  & 1.222\tnote{$\dagger$} & 3.806  & 2.124  & 1.284 \\  \midrule
		\multirow{3}{*}{AR1$\otimes$AR1} & $\hat{\beta}_0$  & 25.103 & 25.223 & 29.266  & 25.033 & 25.032\tnote{$\dagger$} & 27.378 \\
		& $\hat{\beta}_1 (\times 10^4)$ & 16.260 & 6.845  & 130.335 & 16.228 & 6.314\tnote{$\dagger$}  & 112.599  \\
		& $\hat{\beta}_2 (\times 10^8)$  & 8.488  & 3.931  & 61.765  & 7.915  & 3.533\tnote{$\dagger$}  & 54.866  \\  \midrule
		\multirow{3}{*}{\Matern} & $\hat{\beta}_0$ & 21.780 & 23.622 & 18.832  & 21.409 & 23.296 & 15.728\tnote{$\dagger$}  \\
		& $\hat{\beta}_1 (\times 10^4)$ & 11.367 & 5.085  & 122.638 & 6.205  & 2.892\tnote{$\dagger$}  & 88.256 \\
		& $\hat{\beta}_2 (\times 10^8)$  & 5.979  & 2.981  & 60.298  & 3.025  & 1.803\tnote{$\dagger$}  & 43.156  \\ \bottomrule
	\end{tabular}
		    \begin{tablenotes}
	        \item[$\dagger$] \footnotesize Indicates the smallest MSE for the row.
	    \end{tablenotes}
\end{threeparttable}
\end{table}

\subsection{ANOVA}\label{Sec:anova}

Furthermore, ANOVA techniques were used for the analyses of the above results. The analyses were performed for two scenarios of different responses separately. For each response, the coefficients were also accounted for in the model. The objective of the analysis was to investigate the five main factors: two types of design, three bandwidths, three covariance matrices, coefficients $\bm{\beta}$ and the correlation $\epsilon$. The significance patterns of the second order interactions was also of interest. The results are listed in Table \ref{tb:LMMoutput}. 

The results are consistent with what was observed in the subsection above in that, for the linear response, the difference between randomised and systematic designs was not significant. However, for the quadratic response, the design and its interactions with bandwidth and the coefficients were significant. For both scenarios, the correlation intensity and all of its interactions were not significant (Table \ref{tb:LMMoutput}). Hence, GWR performs similarly with either low or high correlation between coefficients. Also of note was that the Bandwidth, and its second order interactions with the variables other than the correlation, was found to be significant for both response types. 

\begin{table}[H]
	\centering
	\caption{ANOVA analyses were conducted on the main factors and their second order interactions. The table lists degrees of freedom (Df), sum of squared errors (Sum Sq) and p-values of F tests (Pr($>$F)). }\label{tb:LMMoutput}
	\begin{tabular}{r | lll | lll}
		\toprule
		 & \multicolumn{3}{c|}{Linear}  & \multicolumn{3}{c}{Quadratic}  \\
    & Df & Sum Sq  & Pr(\textgreater{}F) & Df & Sum Sq  & Pr(\textgreater{}F) \\ \midrule 
Design & 1  & 8.46  & 0.0935  & 1  & 74.03 & \textless{}0.001  \\
Bandwidth & 2  & 7453.09 & \textless{}0.001  & 2  & 90.29  & \textless{}0.001  \\
Covariance ($V_s$)  & 2  & 16958.42  & \textless{}0.001  & 2  & 6122.84 & \textless{}0.001 \\
Coefficients ($\beta$)  & 1  & 903705.52 & \textless{}0.001  & 2  & 1419372.21 & \textless{}0.001  \\
Correlation  ($\epsilon$)  & 1  & 5.63  & 0.171   & 1  & 0.07  & 0.8976  \\
Design:Bandwidth  & 2  & 9.21 & 0.216  & 2  & 113.48   & \textless{}0.001  \\
Design:Covariance & 2  & 3.91 & 0.5217  & 2  & 37.55  & 0.0134 \\
Design:Coefficients  & 1  & 8.45 & 0.0935  & 2  & 147.55 & \textless{}0.001  \\
Design:Correlation  & 1  & 0  & 0.9832  & 1  & 0.01  & 0.9541  \\
Bandwidth:Covariance   & 4  & 12283.13  & \textless{}0.001  & 4  & 3135.47  & \textless{}0.001  \\
Bandwidth:Coefficients & 2  & 7456.81 & \textless{}0.001  & 4  & 179.38 & \textless{}0.001  \\
Bandwidth:Correlation  & 2  & 5.95  & 0.3716  & 2  & 0.85  & 0.9073  \\
Covariance:Coefficients  & 2  & 16959.68  & \textless{}0.001  & 4  & 12248.74 & \textless{}0.001  \\
Covariance:Correlation & 2  & 0.26  & 0.9572  & 2  & 11.57  & 0.2649  \\
Coefficients:Correlation & 1  & 5.63  & 0.171   & 2  & 0.15  & 0.9834  \\ \bottomrule  
	\end{tabular}
\end{table}

% For linear response, the type of design is not a key factor that the design and its interactions with other factors are not significant. On the contrary, if the response is quadratic, design becomes a key factor and the interactions are significant as well. For both scenarios, the correlation intensity and most of its interactions are not significant. Hence, either low or high correlation between coefficients works well with GWR. Moreover, bandwidth selection should be based on the design rather than optimality criteria. The results are consistent with the above boxplots. 

\subsection{AICc Bandwidth selection}\label{Sec:bandselect}

\begin{figure}[H]
	\begin{subfigure}[t]{0.45\textwidth}
		\centering
		\includegraphics[width=\linewidth]{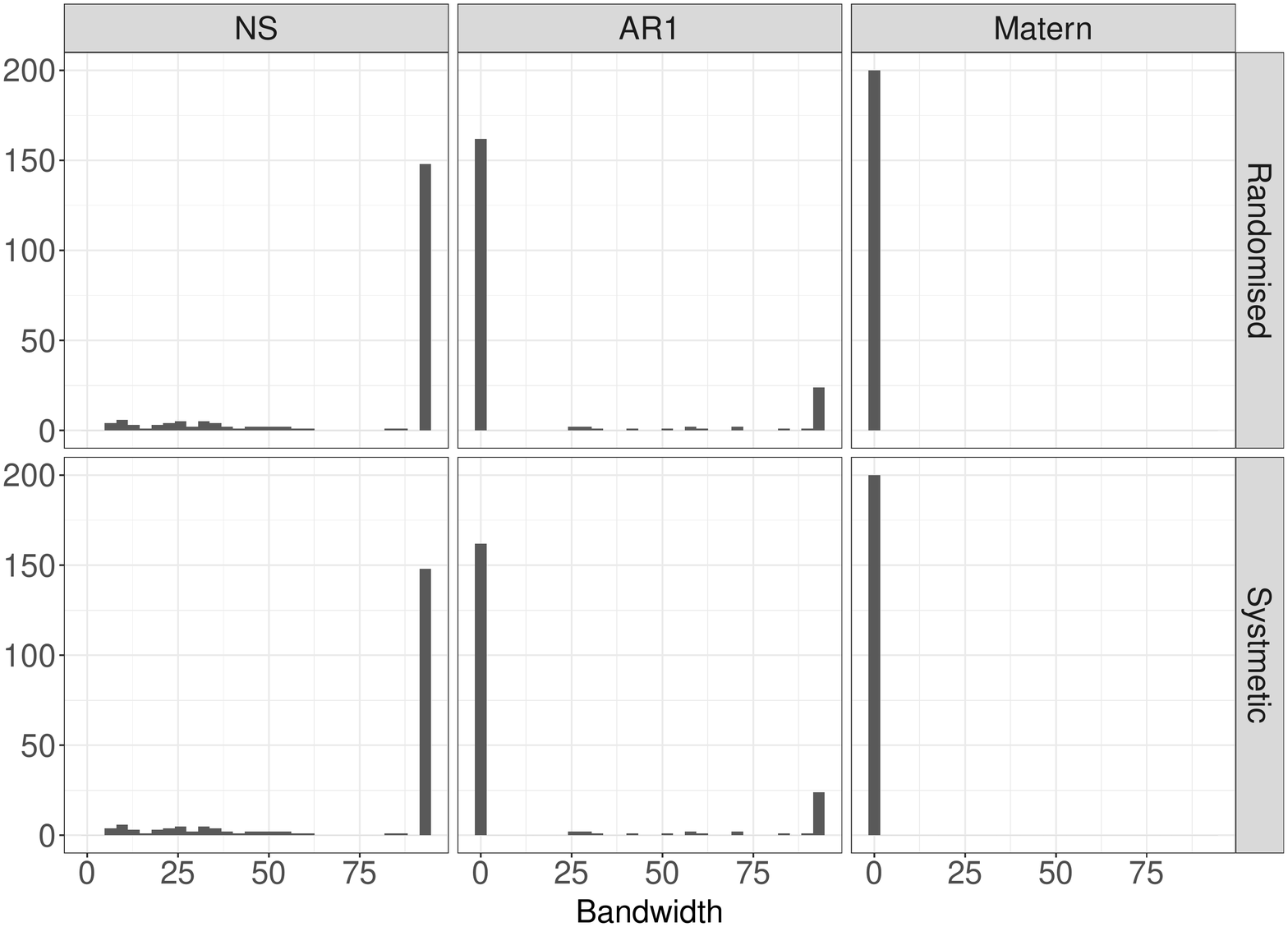}
		\caption{Histogram of optimal bandwidth for linear response.}
	\end{subfigure}
	\hspace{0.05\textwidth}
	\begin{subfigure}[t]{0.45\textwidth}
		\centering
		\includegraphics[width=\linewidth]{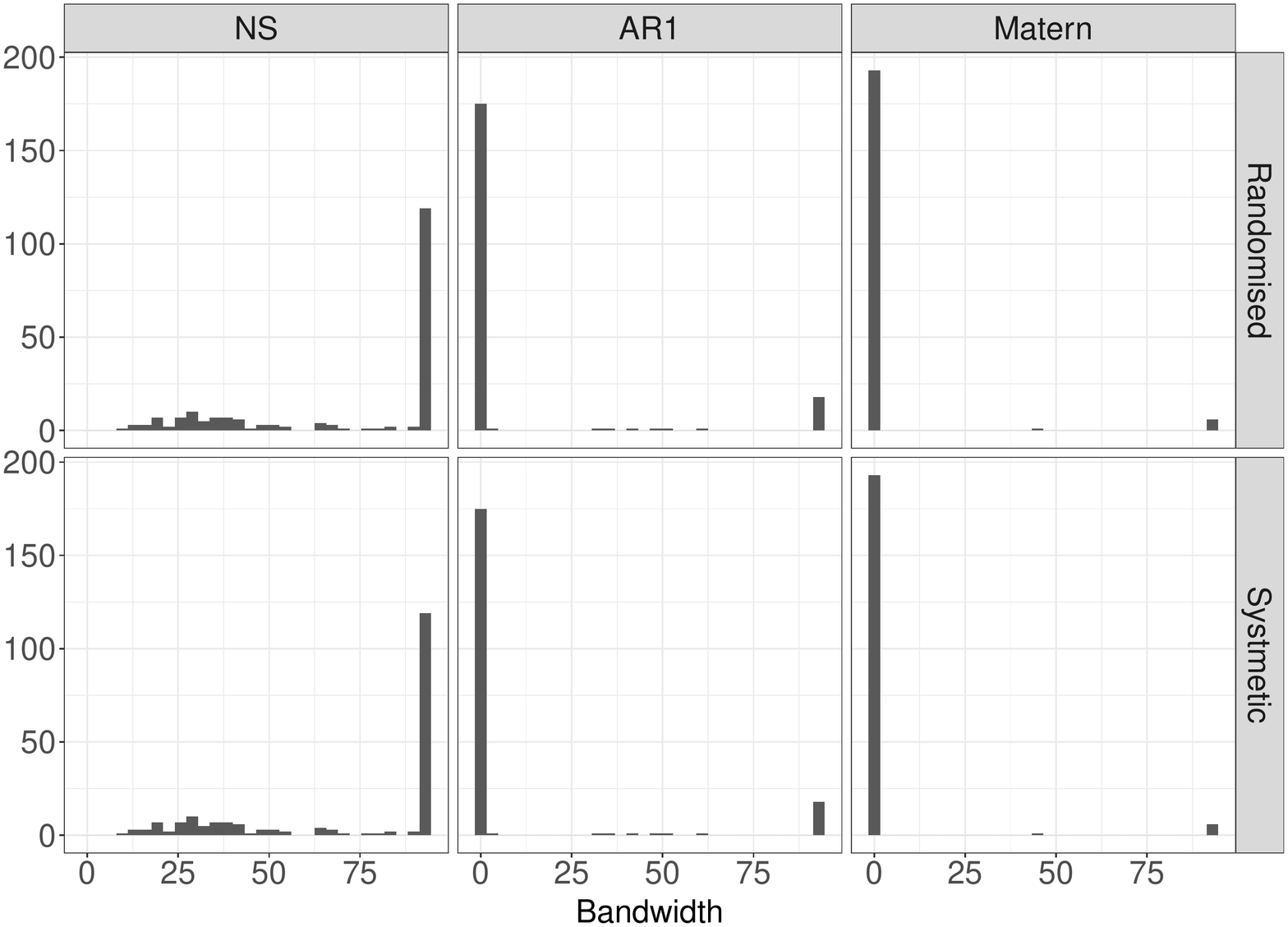}
		\caption{Histogram of optimal bandwidth for quadratic response.}
	\end{subfigure}
	\caption{Histogram of optimal bandwidth found by AICc for linear and quadratic response. }\label{fig:histband}
\end{figure}

From the simulation study, we found that the bandwidth given by AICc skewed to 1 if spatial covariance was included in the model, for all types of the design and nature of the response. If spatial covariance was not introduced, the GWR tended to use all data in one row (Figure \ref{fig:histband}). 

\section{Discussion}\label{Sec:Dis}

%1. Summary of Main Findings
Agronomists and biometricians generally prefer randomised designs for OFE trials. According to the performance metrics used, our simulation study shows a systematic design performed either preferably or similarly to a randomised design for the purposes of creating a varying treatment map. The differentiating factors included primarily the response type and the spatial covariance model, while the correlation amongst the treatment coefficients was not found to be important. These are factors which can be assessed by the farmer beforehand, and this should dictate which design should be used. However, given that a systematic design is easier to implement in the field, and shows little downside for the purposes of creating a varying treatment map, we will advocate for the use of systematic designs. 

%2. How response type affected the performance, and what this means
The response type was the main differentiating factor between randomised and systematic designs. When the response was quadratic, the systematic design performed favourably which contrasted the result for the linear design which saw low differentiation between the designs. Given this, if a farmer expects an approximately linear response in the field, then the selection of the design may not be important. However, given the variable nature of the relationship between response and treatment over a large field (see i.e \textcite{Rakshit2020Novel}), it may be wise to implement a systematic design for the potential outcome of a quadratic relationship. 
%For other more complex response types (i.e polynomials of order greater than 2) it is predicted that a systematic design will also perform favourably. \textcolor{red}{?ref}
%(\textcolor{red}{Might not need a reference)}. 

%2. Discussing Spatial Covariance Structures
%- Some Questions that might be interesting to answer
%- Is a \Matern class more realistic and what are its limitations? %- 
Another consideration for farmers as to which design to use is the expected spatial covariance structure in the field. When no spatial structure was simulated, the differences between the prediction from the systematic and random design were minimal. This result should be expected given if there are no spatial autocorrelations, then the individual query grids are independent observations and therefore the design is not important. However, when a first-order autoregressive structure was simulated the differences were noticeable when a quadratic response was used, showing systematic designs to be preferential. The largest difference between the two designs came when considering the \Matern spatial covariance structure, which showed a clear preference for systematic designs when a quadratic response was considered, and also a small preference for systematic designs for a linear response. Therefore, only if spatial variability was predicted to be negligible in the field would using a randomised design be reasonable given a quadratic response. This assumption of negligible spatial variability would be tough to reason with given the large fields used in on-farm experimentation (add reference), meaning that in application a systematic design should be used. 

% Spatial stationarity is seldom the case in small plot research trials even in highly controlled environments. 

%3. Bandwidth selection and why that is important [good]
There was found to be significant deficiencies in using AICc for bandwidth selection. The AICc-minimising bandwidths skewed to 1 and, in a few cases, ended in 93 (number of rows). Even though the bandwidth was optimal according to AICc, the MSE was higher than when using a fixed bandwidth. Therefore, we recommended using the fixed bandwidth based on the experimental design (5 or 9 in this case), rather than the recommended bandwidth from AICc which is prone to either over-fitting or being too generalised. Selecting the bandwidth based on the experimental design is also theoretically better since only a single measurement is observed in each grid, all levels of the treatment factor should be included in a GWR window at the same time to interpolate the %quadratic curve
relationship. Otherwise, the interpolation is incomplete if more than one level is missing. 

% ... AICc-minimising bandwidth were skewed towards the upper ($sqrt(19^2+92^2)=93.94147$) or lower (1) bounds; two characteristics which are suggestive of underlying stationary or random processes  respectively \textcite{}.  
% In this paper, the simulated data is fitted by GWR, which has been proven to accurately separate treatment from non-treatment influence of yield response. One crucial factor of GWR is bandwidth selection. The AICc-minimising bandwidths skew to 1 and, in a few cases, end in 93 (number of rows). Thus, even though the bandwidth is optimal, the MSE is higher than when using a fixed bandwidth. Therefore, we would use the fixed bandwidth of 5 and 9 based on the experimental design, rather than the recommended bandwidth from AICc. Selecting the bandwidth based on the experimental design is also theoretically better. Since only a single measurement is observed in each grid, all levels of the treatment factor should be included in a GWR window at the same time to interpolate the quadratic curve. Otherwise, the interpolation is incomplete if more than one level is missing. 

%4. Paragraph on Coefficient importance (or the lack thereof)
%The level of correlation between the coefficients as governed by $\epsilon$ was found to no be significant (Table \ref{tb:LMMoutput}). This indicates that GWR performs similarly with either low or high correlation between coefficients. %This is an unexpected/expected result... 

%5. Recommendation of design variations
Given the scope of the paper, some designs and factors were not considered. Designs such as chequerboard or wave designs have been suggested for on-farm experiments \parencite{bramley1999designing}, however weren't able to be considered here. Topographical factors (spatial zones) were also not entertained in our study. Since GWR estimates a global template model and then adjusts it at a local scale across the study region, the variation between zones is ``flushed out'' by the spatial covariance.

%However, this is not practical for farmers as the complex layout of the treatments will increase the cost of sowing and harvesting. Therefore, unless the objective is for variety selection, we do not recommend such designs to agronomists. 

%6. Recommendation of other spatial factors to consider, and why those were ignored. 
%Topographical factors (spatial zones) were also not considered in our study. Since GWR estimates a global template model and then adjusts it at a local scale across the study region, the variation between zones is ``flushed out'' by the spatial covariance.

\section{Conclusion}\label{Sec:Conclusion}

Agronomists and biometricians generally prefer randomised designs for OFE trials. With the purpose of creating a varying treatment map, our simulation study proves that a systematic design produces better performance metrics, under particular circumstances, than a randomised design for large on-farm trials in terms of robustness and smaller MSE on coefficients. On the other hand, if the spatial variation is not considered, or if researchers believe in linear response, either a systematic or a randomised design could be implemented because the difference is not significant. We recommend that, for a large OFE trial with the goal to create a varying treatment map, a systemic design should be used as it has more flexibility in post-experiment statistical modelling. 

% Perhaps we add a comment along the lines - "Further research is needed in the case of OFE (and GWR?) when it comes to more complex designs such as chequerboard, donut etc."

\section{Acknowledgements}

SAGI West gratefully acknowledges the support from the Grains Research and Development Corporation of Australia (GRDC).

\appendix

\section{Figures}

\begin{figure}[!htp]
	\centering
	\includegraphics[width=\linewidth]{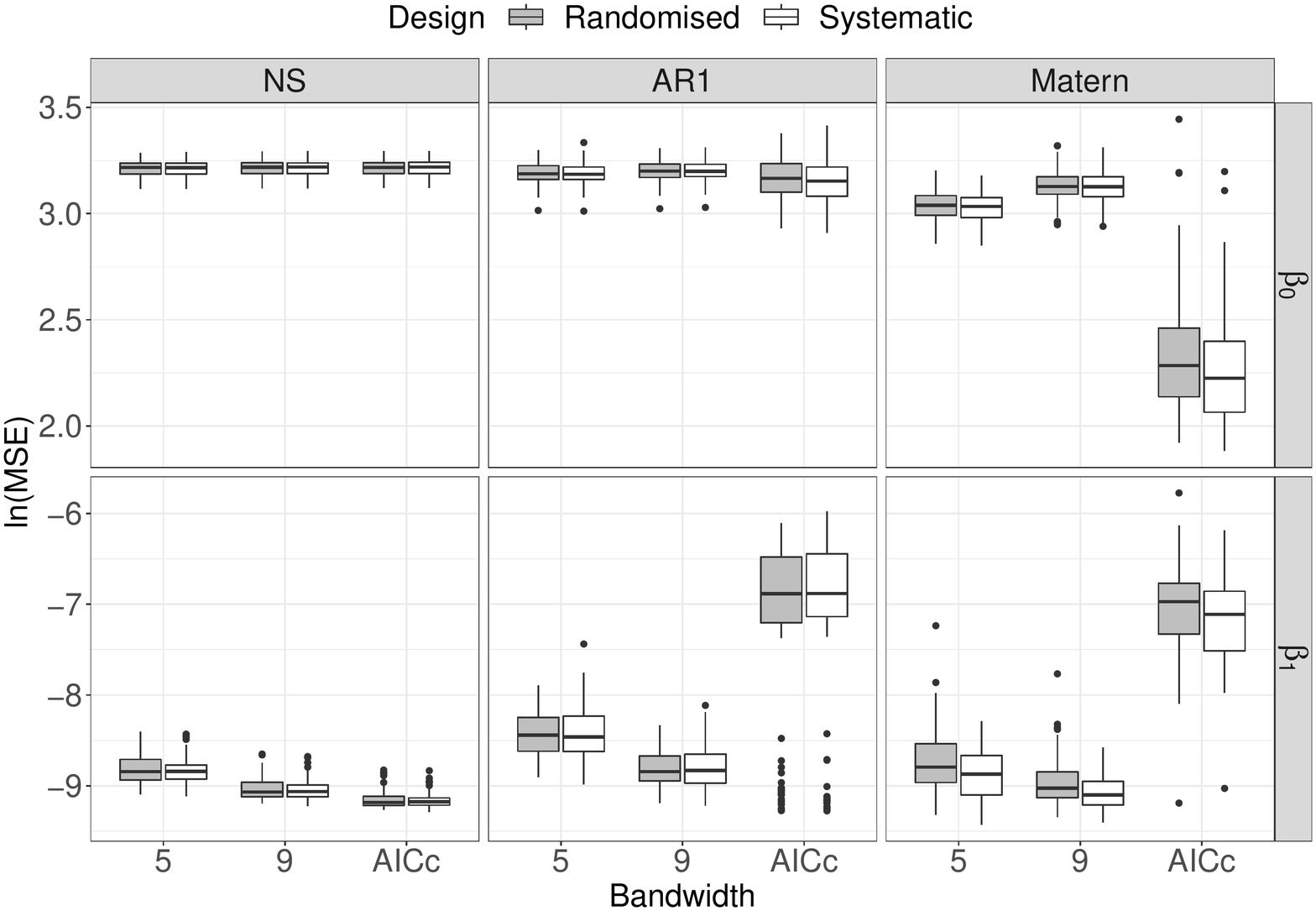}
	\caption{Boxplots of $\ln$(MSE) for $\hat{\beta}_0$ and $\hat{\beta}_1$ in GWR models using different bandwidths for the simulated data with a linear response. The simulated data had different spatial covariance matrices (NS, AR1$\otimes$AR1 and \Matern) and a high correlation between the parameters ($\epsilon=0.1$).}\label{fig:LinBetaMSEeta01}
\end{figure}

\begin{figure}[!htp]
	\centering
	\includegraphics[width=\linewidth]{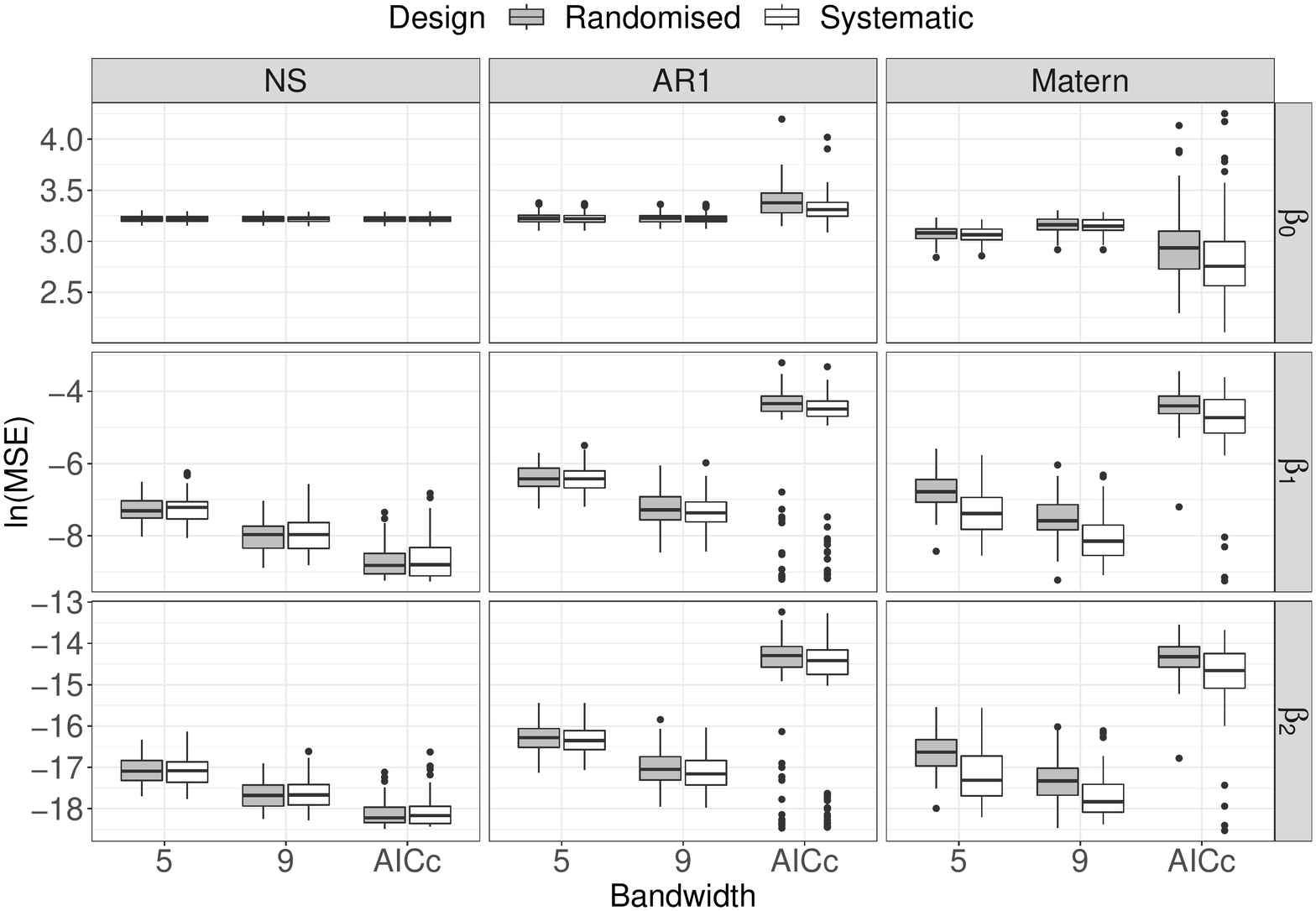}
	\caption{Boxplots of $\ln$(MSE) for $\hat{\beta}_0$, $\hat{\beta}_1$ and $\hat{\beta}_2$ in GWR models using different bandwidths for the simulated data with a quadratic response. The simulated data had different spatial covariance matrices (NS, AR1$\otimes$AR1 and \Matern) and a high correlation amongst the parameters ($\epsilon=0.1$).} \label{fig:QuadBetaMSEeta01}
\end{figure}

\renewcommand\bibname{References}% change bibliography title to references
	%\addcontentsline{toc}{chapter}{Bibliography}
\addtocontents{toc}{Bibliography}
\printbibliography

@book{Abramowitz1974Handbook,
  title = {Handbook of {{Mathematical Functions}}, {{With Formulas}}, {{Graphs}}, and {{Mathematical Tables}}},
  author = {Abramowitz, Milton},
  year = {1974},
  publisher = {{Dover Publications, Inc.}},
  address = {{USA}},
  isbn = {978-0-486-61272-0}
}

@misc{Butler2017ASRemlR,
  title = {{{ASReml}}-{{R Reference Manual Version}} 4},
  author = {Butler, D.G. and Cullis, B.R. and Gilmour, A.R. and Gogel, B.G. and Thompson, R.},
  year = {2017},
  publisher = {{VSN International Ltd, Hemel Hempstead, HP1 1ES, UK.}}
}

@article{Cao2022Bayesian,
title = {Bayesian inference of spatially correlated random parameters for on-farm experiment},
journal = {Field Crops Research},
volume = {281},
pages = {108477},
year = {2022},
issn = {0378-4290},
doi = {https://doi.org/10.1016/j.fcr.2022.108477},
url = {https://www.sciencedirect.com/science/article/pii/S037842902200048X},
author = {Zhanglong Cao and Katia Stefanova and Mark Gibberd and Suman Rakshit}
}

@article{Cook2013Onfarm,
  title = {On-Farm Experimentation},
  author = {Cook, Simon and Cock, James and Oberth{\"u}r, Thomas and Fisher, Myles},
  year = {2013},
  journal = {Better Crop. Plant Food},
  series = {4},
  volume = {97},
  pages = {17--20}
}

@article{Cressie1999Classes,
  title = {Classes of {{Nonseparable}}, {{Spatio}}-{{Temporal Stationary Covariance Functions}}},
  author = {Cressie, Noel and Huang, Hsin-Cheng},
  year = {1999},
  month = dec,
  journal = {Journal of the American Statistical Association},
  volume = {94},
  number = {448},
  pages = {1330--1339},
  publisher = {{Taylor \& Francis}},
  issn = {0162-1459},
  doi = {10.1080/01621459.1999.10473885}
}

@article{Evans2020Assessment,
  title = {Assessment of the {{Use}} of {{Geographically Weighted Regression}} for {{Analysis}} of {{Large On}}-{{Farm Experiments}} and {{Implications}} for {{Practical Application}}},
  author = {Evans, Fiona H. and Recalde Salas, Angela and Rakshit, Suman and Scanlan, Craig A. and Cook, Simon E.},
  year = {2020},
  month = nov,
  journal = {Agronomy},
  volume = {10},
  number = {11},
  pages = {1720},
  publisher = {{Multidisciplinary Digital Publishing Institute}},
  doi = {10.3390/agronomy10111720},
  language = {en}
}

@book{Fisher1934Statistical,
  title = {Statistical Methods for Research Workers},
  author = {Fisher, Ronald Aylmer},
  year = {1934},
  edition = {fifth},
  publisher = {{Oliver and Boyd}},
  address = {{Edinburgh}},
  language = {English}
}

@article{Gollini2015GWmodel,
  title = {{{GWmodel}}: {{An R Package}} for {{Exploring Spatial Heterogeneity Using Geographically Weighted Models}}},
  shorttitle = {{{GWmodel}}},
  author = {Gollini, Isabella and Lu, Binbin and Charlton, Martin and Brunsdon, Christopher and Harris, Paul},
  year = {2015},
  month = feb,
  journal = {Journal of Statistical Software},
  volume = {63},
  number = {1},
  pages = {1--50},
  issn = {1548-7660},
  doi = {10.18637/jss.v063.i17},
  copyright = {Copyright (c) 2013 Isabella Gollini, Binbin Lu, Martin Charlton, Christopher Brunsdon, Paul Harris},
  language = {en}
}

@article{Leung2000Statistical,
  title = {Statistical {{Tests}} for {{Spatial Nonstationarity Based}} on the {{Geographically Weighted Regression Model}}},
  author = {Leung, Yee and Mei, Chang-Lin and Zhang, Wen-Xiu},
  year = {2000},
  month = jan,
  journal = {Environment and Planning A: Economy and Space},
  volume = {32},
  number = {1},
  pages = {9--32},
  publisher = {{SAGE Publications Ltd}},
  issn = {0308-518X},
  doi = {10.1068/a3162}
}

@article{Marchant2019Establishinga,
  title = {Establishing the Precision and Robustness of Farmers' Crop Experiments},
  author = {Marchant, Ben and Rudolph, Sebastian and Roques, Susie and Kindred, Daniel and Gillingham, Vincent and Welham, Sue and Coleman, Colin and {Sylvester-Bradley}, Roger},
  year = {2019},
  month = jan,
  journal = {Field Crops Research},
  volume = {230},
  pages = {31--45},
  issn = {0378-4290},
  doi = {10.1016/j.fcr.2018.10.006},
  language = {en}
}

@book{McElreath2015Statistical,
  title = {Statistical Rethinking: {{A}} Bayesian Course with Examples in {{R}} and Stan},
  author = {McElreath, Richard},
  year = {2015},
  month = dec,
  series = {Chapman and {{Hall}}/{{CRC Texts}} in {{Statistical Science Ser}}.},
  edition = {First},
  volume = {122},
  publisher = {{CRC Press LLC}},
  language = {English}
}

@article{Paez2002General,
  title = {A {{General Framework}} for {{Estimation}} and {{Inference}} of {{Geographically Weighted Regression Models}}: 1. {{Location}}-{{Specific Kernel Bandwidths}} and a {{Test}} for {{Locational Heterogeneity}}},
  shorttitle = {A {{General Framework}} for {{Estimation}} and {{Inference}} of {{Geographically Weighted Regression Models}}},
  author = {P{\'a}ez, Antonio and Uchida, Takashi and Miyamoto, Kazuaki},
  year = {2002},
  month = apr,
  journal = {Environment and Planning A: Economy and Space},
  volume = {34},
  number = {4},
  pages = {733--754},
  publisher = {{SAGE Publications Ltd}},
  issn = {0308-518X},
  doi = {10.1068/a34110}
}

@article{Pandit2019Comparative,
  title = {Comparative Analysis of {{Gaussian Process}} Power Curve Models Based on Different Stationary Covariance Functions for the Purpose of Improving Model Accuracy},
  author = {Pandit, Ravi Kumar and Infield, David},
  year = {2019},
  month = sep,
  journal = {Renewable Energy},
  volume = {140},
  pages = {190--202},
  issn = {0960-1481},
  doi = {10.1016/j.renene.2019.03.047},
  language = {en}
}

@article{Panten2010Enhancing,
  title = {Enhancing the Value of Field Experimentation through Whole-of-Block Designs},
  author = {Panten, K. and Bramley, R. G. V. and Lark, R. M. and Bishop, T. F. A.},
  year = {2010},
  month = apr,
  journal = {Precision Agriculture},
  volume = {11},
  number = {2},
  pages = {198--213},
  issn = {1573-1618},
  doi = {10.1007/s11119-009-9128-y},
  language = {en}
}

@book{Petersen1994Agricultural,
  title = {Agricultural {{Field Experiments}}: {{Design}} and {{Analysis}}},
  author = {Petersen, Roger G},
  year = {1994},
  edition = {1st},
  publisher = {{CRC Press}},
  address = {{Boca Raton}},
  isbn = {978-0-429-07849-1},
  language = {en}
}

@article{Piepho2011Statistical,
  title = {Statistical Aspects of On-Farm Experimentation},
  author = {Piepho, Hans-Peter and Richter, Christel and Spilke, Joachim and Hartung, Karin and Kunick, Arndt},
  year = {2011},
  journal = {Crop \& Pasture Science},
  volume = {62},
  pages = {721--735},
  doi = {10.1071/cp11175}
}

@article{Piepho2013Why,
  title = {Why Randomize Agricultural Experiments?},
  author = {Piepho, H. P. and M{\"o}hring, J. and Williams, E. R.},
  year = {2013},
  journal = {Journal of Agronomy and Crop Science},
  volume = {199},
  number = {5},
  pages = {374--383},
  issn = {09312250},
  doi = {10.1111/jac.12026}
}

@article{Pringle2004FieldScale,
  title = {Field-{{Scale Experiments}} for {{Site}}-{{Specific Crop Management}}. {{Part I}}: {{Design Considerations}}},
  shorttitle = {Field-{{Scale Experiments}} for {{Site}}-{{Specific Crop Management}}. {{Part I}}},
  author = {Pringle, M. J. and Cook, S. E. and McBratney, A. B.},
  year = {2004},
  month = dec,
  journal = {Precision Agriculture},
  volume = {5},
  number = {6},
  pages = {617--624},
  issn = {1573-1618},
  doi = {10.1007/s11119-004-6346-1},
  language = {en}
}

@article{Rakshit2020Novel,
  ids = {Rakshit2020Novela},
  title = {Novel Approach to the Analysis of Spatially-Varying Treatment Effects in on-Farm Experiments},
  author = {Rakshit, Suman and Baddeley, Adrian and Stefanova, Katia and Reeves, Karyn and Chen, Kefei and Cao, Zhanglong and Evans, Fiona and Gibberd, Mark},
  year = {2020},
  journal = {Field Crops Research},
  volume = {255},
  number = {October 2019},
  pages = {107783},
  publisher = {{Elsevier}},
  issn = {0378-4290},
  doi = {10/gg2vv7}
}

@article{Selle2019Flexible,
  title = {Flexible Modelling of Spatial Variation in Agricultural Field Trials with the {{R}} Package {{INLA}}},
  author = {Selle, Maria Lie and Steinsland, Ingelin and Hickey, John M. and Gorjanc, Gregor},
  year = {2019},
  journal = {Theoretical and Applied Genetics},
  volume = {132},
  number = {12},
  pages = {3277--3293},
  publisher = {{Springer Berlin Heidelberg}},
  issn = {14322242},
  doi = {10.1007/s00122-019-03424-y}
}

@article{Verdooren2020History,
  title = {History of the {{Statistical Design}} of {{Agricultural Experiments}}},
  author = {Verdooren, L. Rob},
  year = {2020},
  month = dec,
  journal = {Journal of Agricultural, Biological and Environmental Statistics},
  volume = {25},
  number = {4},
  pages = {457--486},
  issn = {1085-7117, 1537-2693},
  doi = {10.1007/s13253-020-00394-3},
  language = {en}
}

@article{Zimmerman1991Randoma,
  title = {A {{Random Field Approach}} to the {{Analysis}} of {{Field}}-{{Plot Experiments}} and {{Other Spatial Experiments}}},
  author = {Zimmerman, Dale L. and Harville, David A.},
  year = {1991},
  journal = {Biometrics},
  volume = {47},
  number = {1},
  pages = {223--239},
  publisher = {{[Wiley, International Biometric Society]}},
  issn = {0006-341X},
  doi = {10.2307/2532508}
}

@book{Marschner2011,
    address = {San Diego},
    author = {Marschner, Horst},
    doi = {10.1016/C2009-0-63043-9},
    publisher = {San Diego: Elsevier Science \& Technology},
    title = {{Marschner's Mineral Nutrition of Higher Plants}},
    year = {2011}
}

@article{Glynn2007,
    author = {Glynn, Carolyn},
    issn = {0028-646X},
    journal = {New Phytologist},
    number = {3},
    pages = {623--634},
    pmid = {17725548},
    title = {{Testing the Growth-Differentiation Balance Hypothesis: Dynamic Responses of Willows to Nutrient Availability}},
    volume = {176},
    year = {2007},
    doi = {10.1111/j.1469-8137.2007.02203.x}
}

@article{Liben2019,
    author = {Liben, Feyera M and Midega, Tesfaye and Tufa, Tolcha and Wortmann, Charles S},
    doi = {10.1002/agj2.20020},
    journal = {Agronomy Journal},
    title = {{Soil Fertility \& Crop Nutrition Barley and wheat nutrient responses for Shewa, Ethiopia}},
    year = {2019}
}

@article{lu2014gwmodel,
    author = {Binbin Lu and Paul Harris and Martin Charlton and Chris Brunsdon},
    title = {The GWmodel R package: further topics for exploring spatial heterogeneity using geographically weighted models},
    journal = {Geo-spatial Information Science},
    volume = {17},
    number = {2},
    pages = {85-101},
    year  = {2014},
    publisher = {Taylor & Francis},
    doi = {10.1080/10095020.2014.917453},
    URL = {https://doi.org/10.1080/10095020.2014.917453},
    eprint = {https://doi.org/10.1080/10095020.2014.917453}
}

@article{Piepho2018Tutorial,
author = {Piepho, H. P. and Edmondson, R. N.},
title = {A tutorial on the statistical analysis of factorial experiments with qualitative and quantitative treatment factor levels},
journal = {Journal of Agronomy and Crop Science},
volume = {204},
number = {5},
pages = {429-455},
doi = {https://doi.org/10.1111/jac.12267},
url = {https://onlinelibrary.wiley.com/doi/abs/10.1111/jac.12267},
eprint = {https://onlinelibrary.wiley.com/doi/pdf/10.1111/jac.12267},
year = {2018}
}

@article{White2008Agridat,
author = {White, Jeffrey W. and van Evert, Frits K.},
title = {Publishing Agronomic Data},
journal = {Agronomy Journal},
volume = {100},
number = {5},
pages = {1396-1400},
doi = {https://doi.org/10.2134/agronj2008.0080F},
url = {https://acsess.onlinelibrary.wiley.com/doi/abs/10.2134/agronj2008.0080F},
eprint = {https://acsess.onlinelibrary.wiley.com/doi/pdf/10.2134/agronj2008.0080F},
year = {2008}
}

@article{bramley1999designing,
  title={Designing your own on-farm experiments: How precision agriculture can help},
  author={Bramley, RGV and Cook, SE and Adams, ML and Corner, RJ},
  year={1999},
  publisher={Kingston, ACT, Grains Research and Development Corporation}
}
\end{document}